# Shape demagnetization effect on layered magnetoelectric composites

D. A. Pan, J. Lu, Y. Bai, W. Y. Chu and L. J. Qiao[1)]

Environmental Fracture Laboratory of Education of Ministry, Corrosion and Protection Center, University of Science and Technology Beijing, Beijing 100083, P. R. China

**Abstract**: Magnetoelectric (ME) voltage coefficient was always considered independent on in-plane shape of plate layered magnetostrictive-piezoelectric composites due to the oversimplification in ME theoretical models. In this article, we present that in fact the ME voltage coefficient is largely dependent on the in-plane shape owing to shape demagnetization effect on magnetostrictivity. Theoretical analysis and experimental results both indicate that ME voltage coefficient changes notably with the variation of in-plane sizes (length and width) of layered ME properties. ME coefficient increases with the rise of in-plane sizes, and different aspect ratio will also result in different ME coefficient. Proper design of in-plane shape will greatly promote the development of ME devices.

[1)] Corresponding author. Fax: +86 10 6233 2345
*E-mail address*: lqiao@ustb.edu.cn

In the multiferroic materials, the coupling interaction between ferromagnetic and ferroelectric orders could produce some new effects, such as magnetoelectric (ME) or magnetodielectric effect.[1] The ME response, characterized by the appearance of an electric polarization upon applying a magnetic field and/or a magnetization upon applying an electric field, has been observed as an intrinsic effect in some single phase materials.[2, 3] Alternatively, multiferroic composites made by combination of ferromagnetic and ferroelectric substances such as piezoelectric ceramics [e.g., BaTiO$_3$ and lead zirconate titanate (PZT)] and ferrites were found to exhibit large room-temperature extrinsic ME effects recently,[4-7] which has been known as a product property,[8] i.e., a new property of such composites that either individual component phase does not exhibit. This ME effect can be defined as a coupling of magnetic-mechanical-dielectric behavior. That is, when a magnetic field is applied to the composites, the ferromagnetic phase changes the shape magnetostrictivity, and then the strain is passed along to the piezoelectric phase, resulting in an electric polarization.[9] Multiferroic materials have drawn increasing interest due to their multi-functionality, which provides significant potentials for applications in the next-generation multifunctional devices.[10]

Most of previous studies on ME composites always focused on exploring suitable materials and methods to obtain larger ME voltage coefficient $\alpha_{E,33}$ and $\alpha_{E,31}$, but the other transverse ME voltage coefficient $\alpha_{E,32}$ in a rectangular composites was always ignored.[11-19] In addition, most layered ME composite theoretical models stressed on the influence of thickness ratio of ferromagnetic and ferroelectric phases, but paid no attention on the influence of in-plane sizes on ME coefficient.[20-23] In fact, the magnetostrictivity is largely dependent on in-plane sizes due to shape demagnetization effect, so ME coefficients $\alpha_{E,31}$ and $\alpha_{E,32}$ would be different. But no article reported the relationship between ME effect and shape demagnetization effect for plate layered ME composites. Therefore, this article focuses on the effect of in-plane sizes on ME coupling for Ni-PZT-Ni trilayered ME composites.

PZT ceramics were sliced in certain rectangle shapes. The thickness of all PZT samples are 0.8 mm. Then, the sliced samples are materized using autocalytic plating and jointed electrode at both sides. After polarized at 425 K in an electric field of 30-50 kV/cm perpendicular to samples plane, they were bathed in nickel aminosulfonate plating solution, and electrodeposited Ni on both sides. After 8 hours of electro-deposition, the total thicknesses of Ni are about 0.8 mm.

The ME voltage coefficient was measured in a ME measurement systems. The voltage $\delta V$ across the sample was amplified and measured via an oscilloscope. The ME voltage coefficient was calculated based on $\alpha_{E,ij}=\delta V/(t_{PZT} \delta H_j)$, where $t_{PZT}$ is the thickness of PZT, $\delta H_j$ is the amplitude of AC magnetic field generated by Helmholtz coils. In the experiment, the transverse coefficient $\alpha_{E,31}$ and $\alpha_{E,32}$ were measured, as $H_{DC}$ and $\delta H$ were parallel to the length direction and the width direction, respectively. (Fig. 1)

The field dependence and frequency dependence of $\alpha_{E,31}$ and $\alpha_{E,32}$ of a square sample (25×25×0.8 mm$^3$) are shown in Fig. 2. Because the width is equal to the length, $\alpha_{E,31}$ and $\alpha_{E,32}$ exhibit absolutely the same field dependence and frequency dependence. They have the same value at the same bias magnetic field and the same frequency. Both of them reach a maximum at $H_m$=150 Oe, where the maximum are about 43.2 V/cm Oe at the resonance frequencies of 85.4 kHz.

Moreover, the dependence of $\alpha_E$ on $H_{DC}$ and $f$ for a rectangle sample (10×20×0.8 mm$^3$) are shown in Fig. 3. With the rise of $H_{DC}$, both $\alpha_{E,31}$ and $\alpha_{E,32}$ increase rapidly, reach a maximum at $H_m$, then decrease rapidly. However, at the same bias magnetic field, $\alpha_{E,31}$ is always larger than $\alpha_{E,32}$. And $H_m$ of $\alpha_{E,31}$ is 150 Oe but 250 Oe for $\alpha_{E,32}$. In the frequency dependence, the resonant frequencies of $\alpha_{E,31}$ and $\alpha_{E,32}$ are both 87.6 kHz. However, the maximum $\alpha_{E,31}$ of 14.6 V/cm Oe is 3 times larger than the maximum $\alpha_{E,32}$ of 4 V/cm Oe. (Fig. 4b)

Fig. 2 and Fig. 3 show that different in-plane size, shape and direction of applied field will all result in different ME properties. It indicates that ME coupling is not only dependent on raw materials and magnetostrictive-piezoelectric phases thickness ratio, but also on the size and shape of ME composites.

In order to study the effect of size and shape on ME properties, ME composites with different in-plane size and aspect ratio were prepared and measured. Fig. 4 (a) shows the dependence of maximum $\alpha_E$ on the size of rectangular samples with dimension of $\frac{2}{3}L \times L \times 0.8$ mm$^3$, $\frac{1}{2}L \times L \times 0.8$ mm$^3$ and $L \times L \times 0.8$ mm$^3$, respectively, where $L$ is the length of composites. In this figure, only $\alpha_{E,31}$ is shown, because $\alpha_{E,31}$ and $\alpha_{E,32}$ have same variation character, and $\alpha_{E,31}$ is equal to or larger than $\alpha_{E,32}$. With the rise of size (length), the maximum of $\alpha_{E,31}$ increases rapidly.

The difference of ME coupling character originates from the shape demagnetization effect.[24] The ME coefficients are directly proportional to magnetostrictivity $q \sim \delta\lambda/\delta H$, which is sensitive to shape demagnetization. For the ferromagnetic material under a bias magnetic field, lattice will deform along with the spin rotation, and then magnetostriction occurs, which is also sensitive to demagnetization field. For the nonspherical ferromagnet, shape demagnetization is anisotropy, and magnetization is easier along the direction with minimum shape demagnetization energy, so does the magnetostriction.

The shape demagnetization energy is proportional to demagnetization factor $N$.[25,26] The dependence of $N$ on the size for given shapes is shown in Fig. 4(b). The demagnetization factor along the direction of length and width are calculated and signed as $N_1$ and $N_2$. $N_2$ is lager than $N_1$ for a rectangular sample, while $N_1$ is equal to $N_2$ for a square sample. Moreover, $N_1$ and $N_2$ decrease rapidly with the rise of size for both rectangular and square samples.

For a finite sample, if the thicknesses of PZT and Ni are constant, the $\alpha_E$ directly proportional to the magnetostrictivity of Ni. The magnetostrictivity of rectangular samples can be expressed as:[27]

$$q_e = q/(1 + \frac{N}{4\pi}\chi^\sigma) \qquad (1)$$

where $q_e$ is the effective magnetostrictivity, $q$ is the theoretical magnetostrictivity, $N$ is the demagnetization factor and $\chi^\sigma$ is the susceptibility under constant strain condition.

$\alpha_E$ of layered ME composites can be expressed as:[22]

$$\alpha_{E,31} = \frac{-k(q_{11}+q_{21})d_{31}t_{Ni}t_{PZT}}{(s_{11}^{Ni}+s_{12}^{Ni})\varepsilon_{33}kt_{PZT}+(s_{11}^{PZT}+s_{12}^{PZT})\varepsilon_{33}t_{PZT}-2(d_{31})^2 kt_{Ni}} \qquad (2)$$

$$\alpha_{E,32} = \frac{-k(q_{22}+q_{12})d_{31}t_{Ni}t_{PZT}}{(s_{11}^{Ni}+s_{12}^{Ni})\varepsilon_{33}kt_{PZT}+(s_{11}^{PZT}+s_{12}^{PZT})\varepsilon_{33}t_{PZT}-2(d_{31})^2kt_{Ni}} \quad (3)$$

where $k$ is interface coupling parameter, $d_{31}$ is the piezoelectric constant of PZT, $t_{Ni}$ is the thickness of Ni, $t_{PZT}$ is the thickness of PZT, $S$ is the compliance coefficients of Ni or PZT, and $\varepsilon_{33}$ is the dielectric constant of PZT. Because of the influence of demagnetization field in rectangular sample, $q_{11}$ is not equal to $q_{22}$.

For a magnetostrictive material, the magnetostrictivity can expressed as:

$$q_{21} = \beta q_{11}$$
$$q_{22} = \beta q_{12} \quad (4)$$
$$q_{21} = q_{12}$$

where $\beta$ is magnetostrictivity factor of a given shape magnetostrictive material, it donated as the magnetostrictivity ratio of 1 direction and 2 direction and $\beta \leq 1$.

$\chi^\sigma$ is very high for Ni. Combining Eqs. (1) and (4), we arrive at:

$$\beta \approx \sqrt{N_2/N_1} \quad (5)$$

So, we can get:

$$\alpha_{E,31} \propto (1+\beta)/N_1 \quad (6)$$

The calculated $\beta$ for different shapes is shown in table I. Table 1 indicated that magnetostrictive materials have different $\beta$ for different shapes (aspect ration) and have the same $\beta$ for different sizes with the same shape. $\beta$ is increased with the rise of aspect ratio.

Table I. The calculated $\beta$ for different shapes.

|  | 1/2L×L×0.8mm³ | | | 2/3L×L×0.8mm³ | | | L×L×0.8mm³ | | |
| --- | --- | --- | --- | --- | --- | --- | --- | --- | --- |
| $L$(mm) | $N_1$ | $N_2$ | $\beta$ | $N_1$ | $N_2$ | $\beta$ | $N_1$ | $N_2$ | $\beta$ |
| 25 | 0.038 | 0.078 | 0.70 | 0.04 | 0.061 | 0.81 | 0.043 | 0.043 | 1 |
| 20 | 0.045 | 0.092 | 0.70 | 0.047 | 0.072 | 0.81 | 0.05 | 0.05 | 1 |
| 15 | 0.055 | 0.113 | 0.70 | 0.058 | 0.089 | 0.81 | 0.062 | 0.062 | 1 |
| 10 | 0.072 | 0.149 | 0.70 | 0.077 | 0.118 | 0.81 | 0.083 | 0.083 | 1 |
| 5 | 0.112 | 0.23 | 0.70 | 0.121 | 0.185 | 0.81 | 0.133 | 0.133 | 1 |

From equation (6), we can conclude that layered composites have larger ME voltage coefficient when demagnetization factors become smaller with constant thickness of PZT and Ni. For a rectangular sample, $N_2$ is bigger than $N_1$, so $q_{22}$ is smaller than $q_{11}$, that is why the ME voltage coefficient $\alpha_{E,32}$ is smaller than $\alpha_{E,31}$. With the rise of size, the demagnetization factor will decreased, the ME voltage coefficient will increased for rectangular composites with the same shape. Comparing above qualitatively analysis with our results showed in Fig. 2 to Fig. 4, one can find good agreement.

It is worthwhile to note that layered ME composites have different ME voltage coefficient for the plate samples with same ferromagnetic-ferroelectric thickness ratio but different in-plane sizes. The ME coupling is proportional to magnetostrictivity factor $\beta$ and the demagnetization factor $N$. In order to improve the ME coupling, we should make the magnetostrictivity factor higher and higher and make the demagnetization factor lower and lower of ME composites.

Summarily, this article presents the shape demagnetization effect on ME voltage coefficient of the layered ME composites. Different size and shape of ME composites will induce different ME voltage coefficient. With the rise of in-plane size, ME voltage coefficient increases, and different aspect ratio will also result in different ME coefficient. Generally, great attention must be paid to in-plane sizes when one deal with ME models for layered composite or handles ME devices with finite size.

## Acknowledgment

This project was supported by program for Changjiang Scholars, Innovative Research Team in University (IRT 0509) and the National Natural Science Foundation of China under Grant No. 50572006.

**Figure Captions Page**

FIG. 1. Schematic of the rectangular magnetoelectric composites structure.

FIG. 2 (a) Bias magnetic field $H_{DC}$ dependence of $\alpha_{E,31}$ and $\alpha_{E,32}$ for a square sample 25×25×0.8 mm$^3$. (b)Frequency dependence of $\alpha_{E,31}$ and $\alpha_{E,32}$ of the square sample at $H_{DC}=H_m$.

FIG. 3 (a) Bias magnetic field $H_{DC}$ dependence of $\alpha_{E,31}$ and $\alpha_{E,32}$ for a rectangle sample 10×20×0.8 mm$^3$ . The insert show the $H_{DC}$ varied from 0 Oe to 500 Oe. (b) Frequency dependence of $\alpha_{E,31}$ and $\alpha_{E,32}$ at $H_{DC}=H_m$.

FIG. 4. Size dependence of maximum magnetoelectric voltage coefficient $\alpha_{E,31}$ (left) and calculated regularity (right) for $\frac{1}{2}L \times L \times 0.8$ mm$^3$ , $\frac{2}{3}L \times L \times 0.8$ mm$^3$ and $L \times L \times 0.8$ mm$^3$ with the same shape rectangular. (a) Size dependence of calculated demagnetization factor for composites showed in corresponding to (a). (b)

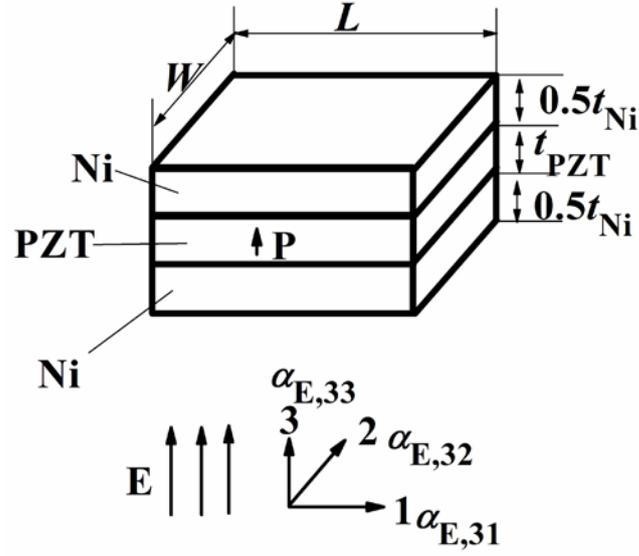

FIG. 1. Schematic of the rectangular magnetoelectric composites structure.

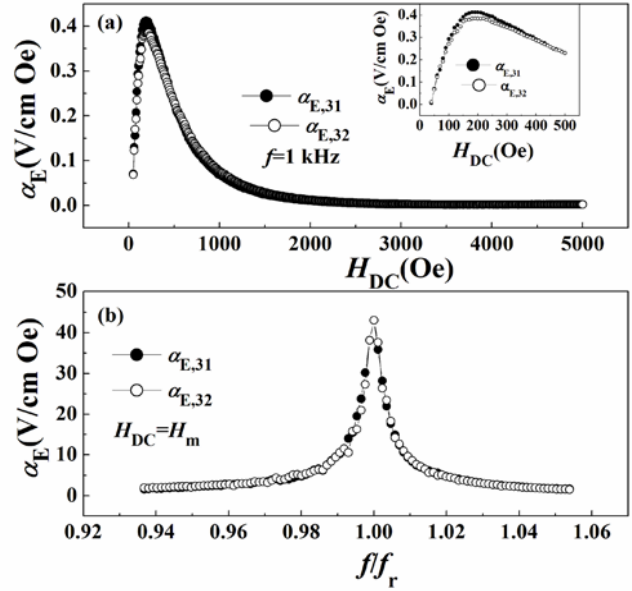

FIG. 2 (a) Bias magnetic field $H_{DC}$ dependence of $\alpha_{E,31}$ and $\alpha_{E,32}$ for a square sample $25\times25\times0.8$ mm$^3$. (b)Frequency dependence of $\alpha_{E,31}$ and $\alpha_{E,32}$ of the square sample at $H_{DC}=H_m$.

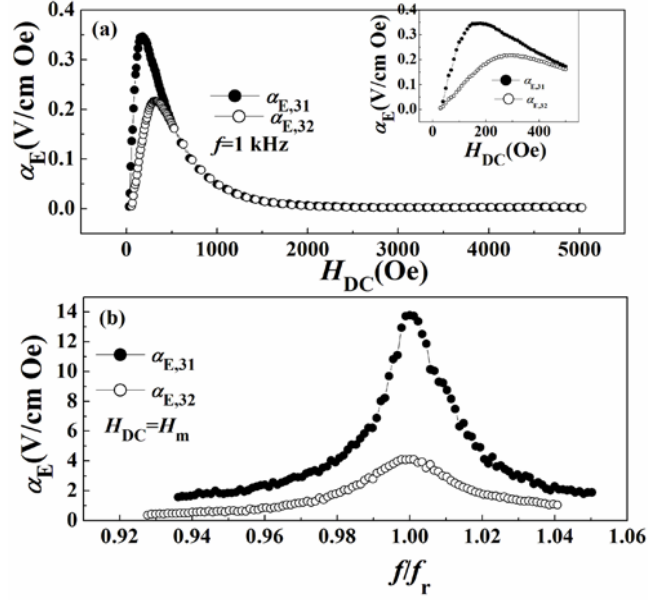

FIG. 3 (a) Bias magnetic field $H_{DC}$ dependence of $\alpha_{E,31}$ and $\alpha_{E,32}$ for a rectangle sample $10\times20\times0.8$ mm$^3$. The insert show the $H_{DC}$ varied from 0 Oe to 500 Oe. (b) Frequency dependence of $\alpha_{E,31}$ and $\alpha_{E,32}$ at $H_{DC}=H_m$.

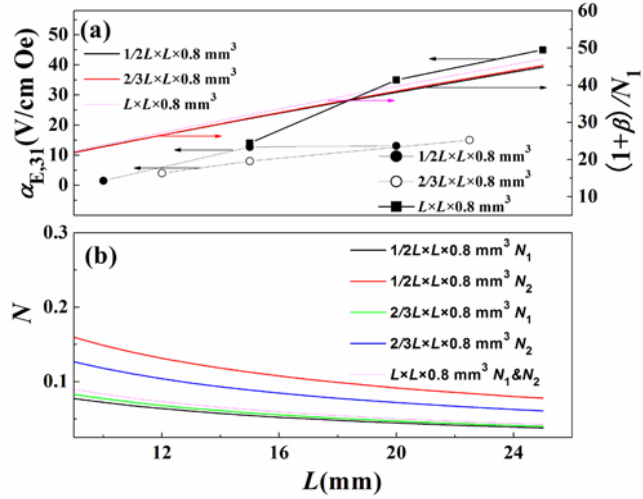

FIG. 4. Size dependence of maximum magnetoelectric voltage coefficient $\alpha_{E,31}$ (left) and calculated regularity (right) for $\frac{1}{2}L\times L\times 0.8$ mm$^3$, $\frac{2}{3}L\times L\times 0.8$ mm$^3$ and $L\times L\times 0.8$ mm$^3$ with the same shape rectangular. (a) Size dependence of calculated demagnetization factor for composites showed in corresponding to (a). (b)